\documentclass[preprint2]{aastex}

\newcommand{\msec}[2]{$#1\mbox{$''\mskip-7.6mu.\,$}#2$}

\newcommand{\myemail}{sdzib@mpifr-bonn.mpg.de}
\newcommand{\UCHII}{\mbox{UC\,H\,{\sc ii}}}
\newcommand{\HII}{\mbox{H\,{\sc ii}}}

\slugcomment{Fourth Draft}

\shorttitle{VLBA Distance to Monoceros}
\shortauthors{Dzib et al.}

\usepackage{epsfig}
\usepackage{longtable}
\usepackage{amssymb}
\usepackage{lscape}
\usepackage{color}

\begin{document}

\newpage

\title{VLBA determination of the distance to nearby star-forming regions\\ VII. Monoceros R2 }

\author{Sergio A. Dzib\altaffilmark{1}, 
Gisela N.\ Ortiz-Le\'on\altaffilmark{2}, 
Laurent Loinard\altaffilmark{1,2}, 
Amy J.\ Mioduszewski\altaffilmark{3},  
Luis F.\ Rodr\'{\i}guez\altaffilmark{2,4},
Rosa M.\ Torres\altaffilmark{5}, 
and
Adam Deller\altaffilmark{6}
}

\altaffiltext{1}{Max Planck Institut f\"ur Radioastronomie, Auf dem H\"ugel 69, 53121 Bonn, Germany (\myemail)}

\altaffiltext{2}{Instituto de Radioastronom\'{\i}a y Astrof\'{\i}sica, Universidad
Nacional Aut\'onoma de M\'exico\\ Apartado Postal 3-72, 58090,
Morelia, Michoac\'an, Mexico }

\altaffiltext{3}{National Radio Astronomy Observatory, Domenici Science Operations Center,\\
1003 Lopezville Road, Socorro, NM 87801, USA}

\altaffiltext{4}{El Colegio Nacional, Donceles 104, 06020, M\'exico, DF, M\'exico}

\altaffiltext{5}{Instituto de Astronom\'{\i}a y Meteorolog\'{\i}a, Universidad de Guadalajara, 
Avenida Vallarta No. 2602, Col. Arcos Vallarta, CP 44130, Guadalajara, Jalisco, M\'exico}

\altaffiltext{6}{ASTRON, The Netherlands Institute for Radio Astronomy, Postbus 2, 7990 AA, 
Dwingeloo, The Netherlands}

\begin{abstract}
We present a series of sixteen Very Long Baseline Array (VLBA) high angular resolution observations
of a cluster of suspected low-mass young stars in the Monoceros R2 region. Four compact and highly 
variable radio sources are detected; three of them in only one epoch, the fourth one a total of seven times. 
This latter source is seen in the direction to the previously known \UCHII\ region VLA~1, and has 
radio properties that resemble those of magnetically active stars; we shall call it VLA~1$^\star$.
We model its displacement on the celestial sphere as a combination of proper motion and trigonometric 
parallax. The fit obtained using a uniform proper motion yields a parallax $\varpi$ = 1.10 $\pm$ 0.18 mas, 
but with a fairly high post-fit dispersion. If acceleration terms 
(probably due to an undetected companion) are included, the quality of the fit 
improves dramatically, and the best estimate of the parallax becomes $\varpi$ = 1.12 $\pm$ 0.05 mas. 
The magnitude of the fitted acceleration suggest an orbital period of order a decade.
The measured parallax corresponds to a distance $d$ = 893$^{+44}_{-40}$ pc, in very good agreement with previous, indirect, determinations.
\end{abstract}

\keywords{astrometry --- radio continuum: stars ---  radiation mechanisms: non--thermal  --- techniques: interferometric}

\section{Introduction}

Very Long Baseline Interferometry (VLBI) at radio frequencies is an observing technique that provides 
images with very high angular resolution (of order 1 mas at $\nu$ $\sim$ 10 GHz) and very accurate 
astrometric registration relative to distant quasars (typically better than 0.1 mas, even in moderate 
signal-to-noise situations; e.g.\ Reid \& Honma 2014). An interesting application of these characteristics
is the direct determination of distances through the accurate measurement of trigonometric parallaxes using
multi-epoch VLBI observations. Magnetically active Young Stellar Objects (YSOs) have been a target of choice
because they are associated with bright and compact radio sources that can easily be detected in VLBI observations
(e.g.\ Loinard et al.\ 2005, 2007, 2008; Menten et al.\ 2007; Torres et al.\ 2007, 2009, 2012; Dzib et al.\ 2010,
2011; Ortiz-Le\'on et al.\ 2016, in preparation). VLBI observations are particularly appropriate for the case of young stars (e.g., 
Loinard et al.\ 2011; Brunthaler et al.\ 2011) 
because they are often highly embedded inside of their parental cloud, and therefore highly obscured. As a consequence, 
optical and near-IR telescopes (including the Gaia astrometry mission) are at a disadvantage for this type of sources.
In this paper, we will present  multi-epoch VLBI observations of a suspected cluster of magnetically active low-mass
YSOs in the Monoceros R2 high-mass star-forming region.

The Monoceros R2  region was initially identified as a chain of reflection nebulae illuminated by A-- and
B--type stars (Gyulbudaghian et al.\ 1978 and references therein) and named GGD 11 to 17. Those nebulae are
associated with a giant molecular cloud that harbors numerous other signposts of active massive star-formation
(water masers, bright infrared sources, etc.). Indeed, one of the first molecular outflows ever reported 
(Rodr\'{\i}guez et al.\ 1982) is located in the GGD 12--15 sub-region of Monoceros. In this very same region,
G\'omez et al.\ (2000, 2002) discovered a cluster of nine compact radio sources surrounding a cometary Ultra 
Compact \HII\ region (\UCHII), and distributed over an area of less than 400 arcsec$^2$. The \UCHII\ region was
named VLA~1 and the compact sources VLA~2 to VLA~10. G\'omez et 
al.~(2000, 2002) argued that most of these 
compact radio sources are associated with magnetically active low-mass YSOs because of their variability and 
spectral indices. As a consequence, they are 
good candidates for VLBI observations.

Obtaining an accurate distance to Monoceros is of high importance since it is, after Orion at 414$\,\pm\,$7 pc (Menten et al.\ 2007) 
and Cepheus A at 700$\,\pm\,$30 pc (Moscadelli et al.\ 2009; Dzib et al.\ 2011), one of the nearest regions where high-mass star 
formation is occurring. 
Distances to Monoceros R2 have been estimated mainly using spectroscopic and photometric studies. The first
measurements are from Racine (1968), who estimated a distance of 830$\,\pm\,$50 pc; and Rozhkovski \& Kurchakov~(1968)
who found a distance of 700 pc (see also Downes et al.\ 1975). Later on, Racine \& Van der Bergh (1970), 
in a proceeding conference, gave a revised distance of 950 pc, without further details. In the most recent paper 
using that technique, Herbst \& Racine (1976) reported 830$\,\pm\,$50 pc; this last value is the most commonly 
adopted distance to Monoceros R2 in the astronomical literature (Carpenter \& Hodapp 2008). On the other hand, Rodr\'{\i}guez 
et al.\ (1980) use CO line observations to derive a kinematic distance of 1 kpc, which is another commonly used 
distance for Monoceros R2.  Finally, Reid et al. (2016) have made available a bayesian distance calculator 
that improves the determination of the kinematic distances to sources in the galactic spiral arms. By using this tool and 
a $v_{lsr}$ = +11 $\pm$ 1 km s$^{-1}$ for GGD 12--15 (Rodr\'{\i}guez et al. 1980) we obtained two possible distances.
The most likely distance, with a probability density of 17 kpc$^{-1}$, is 760$\,\pm\,70$ pc that is in agreement with 
the distance obtained by Herbst \& Racine (1976). The second possible distance, with a probability density of just 5 kpc$^{-1}$,
is around 420 pc.

\begin{figure}[!ht]
\begin{center}
\includegraphics[width=1.0\linewidth,trim= 30 0 40 40, clip]{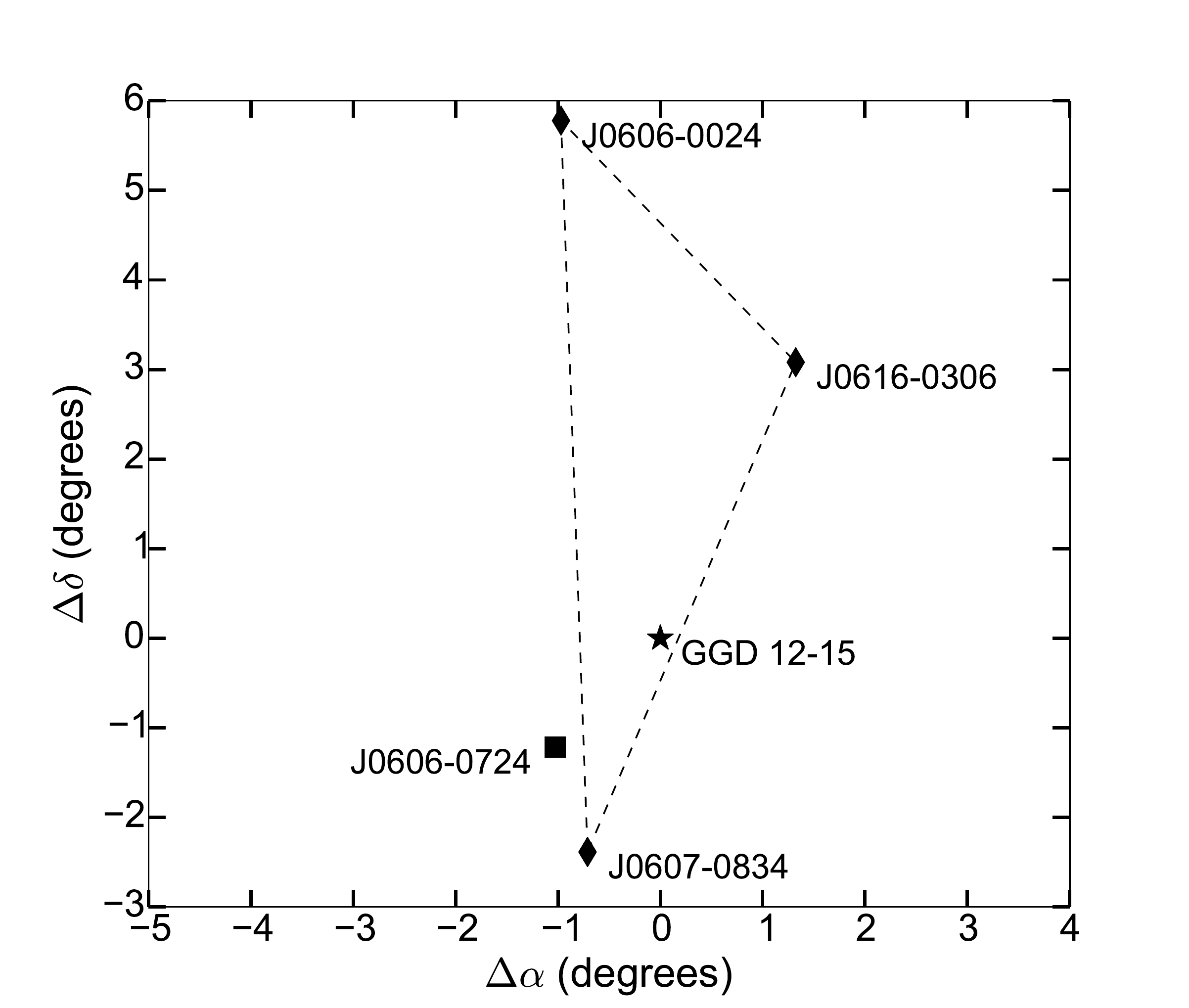}
\end{center}
\caption{Positions of the main (black square) and secondary calibrators (black diamonds) relative to the target
position (black star).}
\label{fig:calpos}
\end{figure}

\section{Observations}

\begin{figure*}[!ht]
\begin{center}
\includegraphics[width=1.0\linewidth]{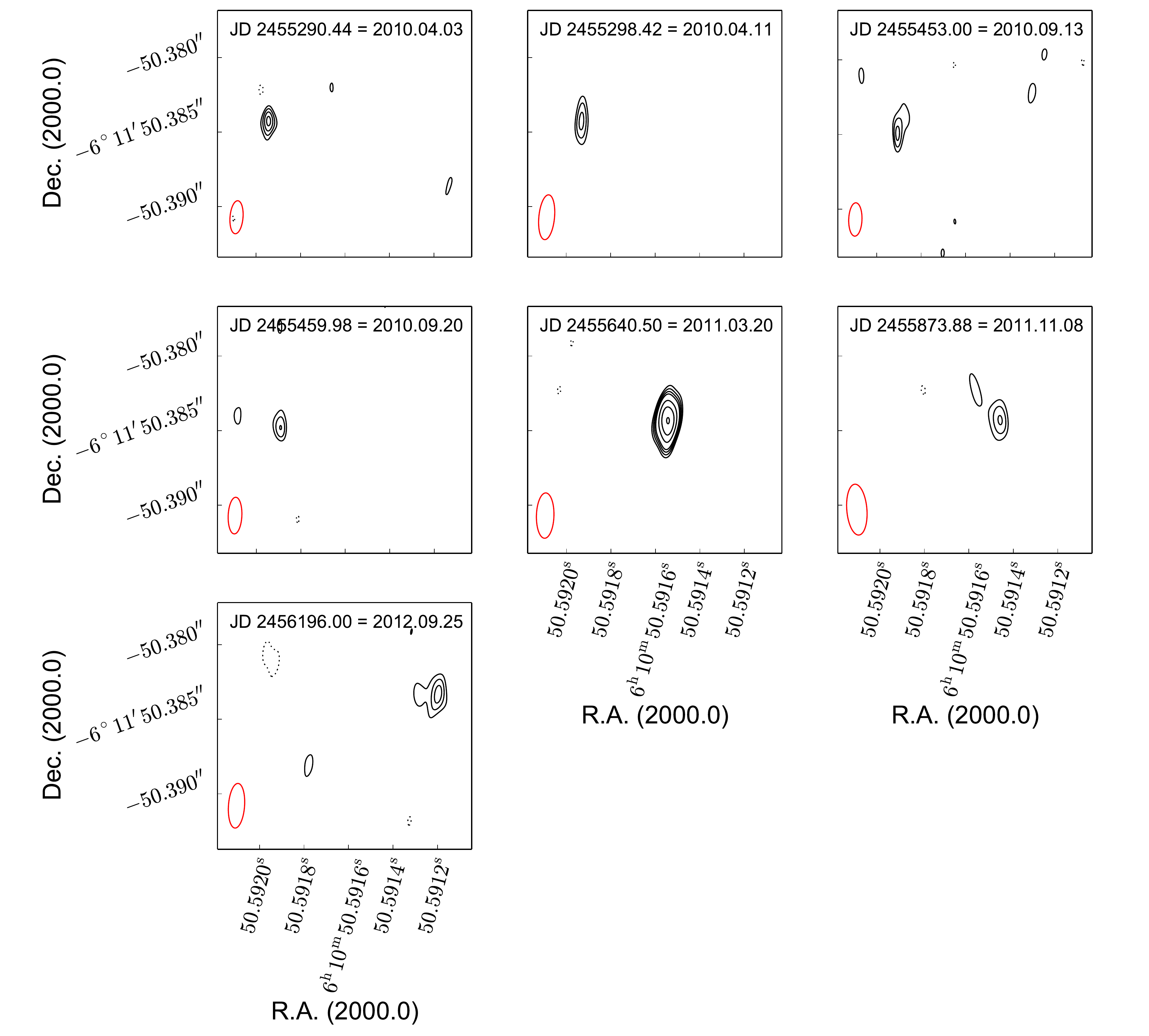}
\end{center}
\caption{Images of VLA~1$^\star$ at the six detected epochs. The synthesized beam is displayed
as a red ellipse at the bottom left of each panel. The contour levels are -3, 3, 4, 5, 6, 9, 12, and 15
times the noise level in each image, which are given in Table \ref{tab:rpfi}.}
\label{fig:VLA1det}
\end{figure*}

We observed the 10 VLA sources from G\'omez et al.\ (2000, 2002) using the VLBI technique. Specifically,
the observations were collected using the Very Long Baseline Array (VLBA; Napier et al. 1994) operated
in combination with the 100-m Robert C. Byrd Green Bank telescope (GBT; Prestage et al.\ 2009) at a wavelength of 3.6~cm 
($\nu$ = 8.42 {\rm GHz}). A total of 17 observations were performed (see Table \ref{tab:rpfi}). 
We first undertook a pilot study (BL169) which consisted of three epochs, in which we detected three sources.
Following the successful detection of three of the target sources, we initiated a series of multi-epoch 
observations starting in September 2010, as part of the project BL176. The observations, including the 
pilot epochs, were accommodated close to the equinoxes when the maximum elongation of the parallactic
ellipse occurs (specifically for Monoceros, the maximum elongation occurs on March and September 24 of 
each year). During each equinox a maximum of three observations were scheduled separated 
from one another by a few days to a few weeks. As the sources are highly variable, this observing mode 
increases the probability of obtaining at least one detection during each equinox. The last observation
was collected in October 2012. 

The spectral setups were different for projects BL169 and BL176. In the former, four base-band channels 
(BBCs) were observed. Each BBCs contained 128 channels of 0.125 MHz each. This setup was chosen
to enable the mapping of the entire area containing our 10 targets without significant bandwidth smearing
loss. The BL176 observations, on the other hand, used only 16 channels of 1~MHz each, but took advantage 
of the (then, newly available) multi-phase center capability  provided by the VLBA DifX digital correlator
(Deller et al.\ 2011). In this observing mode, a different phase center was positioned at the location of 
each one of our 10 targets, generating 10 separate and smaller visibility datasets.

The main phase calibrator for all epochs was J0606--0724, located at an angular distance of 1.6 degrees 
from the targets. To improve the quality of phase calibration, we also observed the secondary calibrators 
J0607--0834, J0616--0306 and J0606--0024, which are distributed around our target sources as shown in 
Figure \ref{fig:calpos}. Additionally, about two dozen ICRF quasars spread over the entire visible sky were 
observed periodically during the observations (conforming so-called {\em geodetic blocks}); those are used 
to improve tropospheric calibration (e.g., Reid \& Brunthaler 2004). The observations obtained as part of 
project BL169 had a total duration of five hours while those corresponding to project BL176 had a duration 
of nine hours. Each individual observation contained three 30-minute geodetic blocks, scheduled at the
beginning, the middle, and the end of each observation. The observations of the targets were recorded as 
part of cycles with two minutes spent on-source, and one minute 
spent on the main phase calibrator. Roughly every 30 minutes, we also observed the secondary calibrators,
spending one minute on each one. 

Two of the observations were affected by logistical issues and were excluded from our analysis: Epoch
BL176l was a total loss, with no usable data recorded, while epoch BL176h was affected by a mistake during
correlation which led to no dataset being produced for the target phase centers.
Additionally, in the epoch BL176f the setup used to observe the ICRF quasars was erroneously kept until 
half-way through the observations, meaning the first half of the target data was unusable. This limited 
the observing time, and explains the higher noise level 
recorder in that observation (Table \ref{tab:rpfi}). In epoch BL176m the Tsys information for the GBT 
antenna was not generated, and it was artificially created a posteriori using the Tsys information from 
the epoch BL176c that was taken almost on the same date but two years before. In this case we down-weighted
the visibilities containing GBT data to minimize the impact of the resulting poorer amplitude calibration.
We note that the calibrator data are self-calibrated in amplitude as part of the data processing.
This helped to mitigate the lack of Tsys measurements for the GBT.
Finally for epochs BL176d and BL176e the field related to source VLA 9 was not correlated. 

The data were edited and calibrated using the Astronomical Image Processing System (AIPS; Greisen 2003). 
The basic data reduction followed the standard VLBA procedure for phase-referenced observations, including 
the multi-calibrator schemes and the tropospheric and clock corrections obtained from the geodetic blocks.
These calibrations were described in detail by Loinard et al.\ (2007), Torres et al.\ (2007) and Dzib et al.\ (2010). 
After calibration, the visibilities were first imaged with a pixel size of 50 $\mu$as using a natural weighting 
scheme\footnote{ The weigthing schemes are derived and discussed by Briggs (1995).
Briefly, the ROBUST parameters specifies how the visibilities are weighted before imaging.
With ROBUST = +5 (natural weighting), the visibilities are weighted according to their density in the $(u,v)$ plane. 
This leads to the best possible noise level, but since the central portion of the $(u,v)$ plane tends to be more 
densely populated than the outer parts, natural weighting produces the lowest possible angular resolution. 
ROBUST = $-$5 (uniform weighting), on the other, produces a uniform weighting of the visibilities across 
the $(u,v)$ plane. This results in the highest possible resolution, but worst possible noise level. Intermediate 
values (in AIPS, ROBUST varies continuously between the $\pm$ 5 extrema) correspond to intermediate weighting 
schemes between natural and uniform. In many cases, ROBUST values around zero provide a good compromise yielding 
a substantial gain in resolution over natural weighting, at the cost of only a minor loss in sensitivity.} 
(ROBUST = 5 in AIPS). As this scheme provides the best possible noise levels, we used these images 
to search for source detections. When a detection was obtained, we constructed new images with a weighting
scheme intermediate between natural and uniform (ROBUST = 0 or ROBUST = 2, depending on the intensity of 
the detection). This enabled
a slighty better determination of the source 
positions at each epoch. The r.m.s.\ noise levels in the final images were 18 -- 41 $\mu$Jy beam$^{-1}$. 
The parameters of images in individual epochs are given in Table \ref{tab:rpfi}. From these images, the 
source position, when detected, was determined by using a two-dimensional fitting procedure (task JMFIT in AIPS).
The brightness temperature was calculated following:
$$
T_b[{\rm K}]\simeq6.26\times10^{6}\left(\frac{S_\nu}{\rm mJy}\right)\left(\frac{\theta_{\rm maj}}{\rm mas}\right)^{-1}\left(\frac{\theta_{\rm min}}{\rm mas}\right)^{-1},
$$
where $\theta_{\rm maj}$ and $\theta_{\rm min}$ are the semi-major and semi-minor axis of the deconvolved 
source size determined by JMFIT.

\begin{table}[!h]
\scriptsize
\tablewidth{\linewidth}
\centering
\caption{Observation summary.}
\begin{tabular}{lccc}\hline\hline
       &UT date      & Synthesized beam                                  & $\sigma_{\rm noise}$  \\
Project\tablenotemark{a}&(yyyy.mm.dd) &  ($\theta_{\rm maj}\times\theta_{\rm min};$ P.A.) & ($\mu$Jy$\,$bm$^{-1}$)\\%
\hline\noalign{\smallskip}
BL169a&2010.03.28&\msec{0}{0034}$\times$\msec{0}{0010};\,-6.0$^\circ$&25\\
BL169b&2010.04.03&\msec{0}{0033}$\times$\msec{0}{0010};\,-6.4$^\circ$&23\\
\hspace{0.4cm}(R0)&      &\msec{0}{0022}$\times$\msec{0}{0009};\,-4.7$^\circ$&35\\
BL169c&2010.04.11&\msec{0}{0036}$\times$\msec{0}{0011};\,-6.3$^\circ$&24\\
\hspace{0.4cm}(R2)&      &\msec{0}{0030}$\times$\msec{0}{0011};\,-5.4$^\circ$&27\\
BL176a&2010.09.13&\msec{0}{0033}$\times$\msec{0}{0011};\,-5.1$^\circ$&18\\
\hspace{0.4cm}(R0)& &\msec{0}{0022}$\times$\msec{0}{0009};\,-2.0$^\circ$&25\\
BL176b&2010.09.20&\msec{0}{0033}$\times$\msec{0}{0011};\,-5.2$^\circ$&19\\
\hspace{0.4cm}(R0)& &\msec{0}{0025}$\times$\msec{0}{0009};\,-2.5$^\circ$&26\\
BL176c&2010.09.26&\msec{0}{0033}$\times$\msec{0}{0011};\,-4.3$^\circ$&20\\
BL176d&2011.03.20&\msec{0}{0035}$\times$\msec{0}{0012};\,-3.0$^\circ$&18\\
\hspace{0.4cm}(R0)& &\msec{0}{0023}$\times$\msec{0}{0010};\,+0.2$^\circ$&26\\
BL176e&2011.03.25&\msec{0}{0037}$\times$\msec{0}{0011};\,-4.4$^\circ$&20\\
BL176f&2011.09.15&\msec{0}{0025}$\times$\msec{0}{0011};\,+3.9$^\circ$&41\\
BL176g&2011.09.19&\msec{0}{0057}$\times$\msec{0}{0017};\,-12.0$^\circ$&24\\
BL176h&2011.09.26&\nodata&\nodata\\
BL176i&2011.11.06&\msec{0}{0037}$\times$\msec{0}{0012};\,-3.8$^\circ$&18\\
BL176j&2011.11.08&\msec{0}{0036}$\times$\msec{0}{0013};\,0.5$^\circ$&20\\
BL176k&2012.04.01&\msec{0}{0033}$\times$\msec{0}{0012};\,-4.2$^\circ$&18\\
BL176l&2012.09.20&\nodata&\nodata\\
BL176m&2012.09.25&\msec{0}{0030}$\times$\msec{0}{0011};\,-4.1$^\circ$&18\\
BL176n&2012.10.01&\msec{0}{0036}$\times$\msec{0}{0011};\,-3.9$^\circ$&17\\
\hline\hline
\tablenotetext{a}{The first line for each observation shows the parameters of the images obtained with 
natural weighting (ROBUST = 5). When a second line is present, it contains information on images reconstructed
for the same epoch, but with a different weighting scheme: R0 and R2 stand for ROBUST = 0 and ROBUST = 2, respectively (see text). }

\label{tab:rpfi}
\end{tabular}
\end{table}

\begin{table*}[!htbp]
\scriptsize
\centering
\caption{Calendar and Julian dates, measured source positions, flux densities, noise levels, and brightness temperatures of detections.}
\begin{tabular}{lcccccccc}\hline\hline
Mean UT date &Julian Day & $\alpha$(J2000.0) & $\sigma_{\alpha}$ & $\delta$(J2000.0) & $\sigma_\delta$ & $f_\nu$ &T$_b$\\
(yyyy.mm.dd/hh:mm)& & $06^{\rm h}10^{{\rm m}}$& & $-06^{\circ}$& & (mJy)&  ($10^6$ K)\\
\hline\noalign{\smallskip}
\multicolumn{8}{l}{VLA~1}\\
2010.04.03/22:33&2455290.44&50\rlap.{$^{\rm s}$}591943 &0\rlap.{$^{\rm s}$}0000048&$11'$ 50\rlap.{''}38440&0\rlap.{''}00013&0.23$\pm$0.04\tablenotemark{a}&$>$3.0\tablenotemark{c} \\
2010.04.11/22:04&2455298.42&50\rlap.{$^{\rm s}$}591930 &0\rlap.{$^{\rm s}$}0000045&$11'$ 50\rlap.{''}38429&0\rlap.{''}00027&0.15$\pm$0.03\tablenotemark{b}&$>$1.9\tablenotemark{c} \\
2010.09.13/12:00&2455453.00&50\rlap.{$^{\rm s}$}591896 &0\rlap.{$^{\rm s}$}0000053&$11'$ 50\rlap.{''}38454&0\rlap.{''}00026&0.17$\pm$0.05\tablenotemark{a}&$>$2.2\tablenotemark{c} \\
2010.09.20/11:31&2455459.98&50\rlap.{$^{\rm s}$}591893 &0\rlap.{$^{\rm s}$}0000058&$11'$ 50\rlap.{''}38463&0\rlap.{''}00020&0.23$\pm$0.03\tablenotemark{a}&$>$2.5\tablenotemark{c} \\
2011.03.20/00:00&2455640.50&50\rlap.{$^{\rm s}$}591548 &0\rlap.{$^{\rm s}$}0000023&$11'$ 50\rlap.{''}38434&0\rlap.{''}00008&0.45$\pm$0.03\tablenotemark{a}&$>$2.0\tablenotemark{d} \\
2011.11.08/09:07&2455873.88&50\rlap.{$^{\rm s}$}591464 &0\rlap.{$^{\rm s}$}0000081&$11'$ 50\rlap.{''}38442&0\rlap.{''}00026&0.10$\pm$0.03&$>$0.5\tablenotemark{c} \\
2012.09.25/12:03&2456196.00&50\rlap.{$^{\rm s}$}591200 &0\rlap.{$^{\rm s}$}0000072&$11'$ 50\rlap.{''}38337&0\rlap.{''}00024&0.10$\pm$0.02&$>$0.2\tablenotemark{c} \\
\noalign{\smallskip}
\multicolumn{8}{l}{VLA 3}\\
2010.03.28/23:00&2455284.46&49\rlap.{$^{\rm s}$}945410 &0\rlap.{$^{\rm s}$}0000029&$11'$ 45\rlap.{''}82133&0\rlap.{''}00012&0.26$\pm$0.03&$>$1.9\tablenotemark{c} \\
\noalign{\smallskip}
\multicolumn{8}{l}{VLA 8}\\
2011.03.26/01:15&2455646.55&49\rlap.{$^{\rm s}$}212366 &0\rlap.{$^{\rm s}$}0000033&$11'$ 29\rlap.{''}93380&0\rlap.{''}00015&0.28$\pm$0.04&$>$0.5\tablenotemark{d} \\
\noalign{\smallskip}
\multicolumn{8}{l}{VLA 10}\\
2010.04.11/22:04&2455298.42&50\rlap.{$^{\rm s}$}504350 &0\rlap.{$^{\rm s}$}0000051&$12'$ 04\rlap.{''}21946&0\rlap.{''}00023&0.15$\pm$0.03&$>$1.2\tablenotemark{c} \\
\hline\hline
\tablenotetext{a}{From image produced with ROBUST=0.}
\tablenotetext{b}{From image produced with ROBUST=2.}
\tablenotetext{c}{Source is unresolved, we used the size of the synthesized beam as the nominal size.}
\tablenotetext{d}{Source can be resolved but can also be a point source, we used the maximum fitted size 
to place a lower limit on the brightness temperature.}
\label{tab:hri}
\end{tabular}
\end{table*}

\section{Results}

A total of four distinct sources were detected in our observations. A source associated with VLA~1
was detected seven times (Figure \ref{fig:VLA1det}). Additionally, we detected a source associated
with VLA~3 at 10$\sigma$, a source associated with VLA~8 at 11$\sigma$, and a source associated with
VLA~10 at 6$\sigma$. All three of these cases, however, were single-epoch detections. The parameters
of all detections are given in Table \ref{tab:hri}. { In most cases, the deconvolved size for 
the detected sources was consistent with a point-like structure.  In two cases the source could be
modeled with a slightly resolved source, although a point-like structure cannot be discarded.} Given the
measured flux densities and the 
compactness of the emission, each of the detections implies brightness temperatures in excess of
$\sim$10$^6$~K (Table \ref{tab:hri}). In addition, the sources appear to be highly variable: the
VLBA source associated with VLA~1 shows variations by a factor of nearly 5, while the sources 
associated with VLA~3, 8, and 10 should have been detected at most epochs if their fluxes remained 
steady. The fact that they were not suggests that they go through occasional flaring activity when
their flux increases at least by a factor of 
several. This combination of high brightness temperature and high variability strongly suggests
that the detected sources are flaring low-mass stars with coronal emission. This was previously
suggested for VLA~3, VLA~8 and VLA~10 by Gom\'ez et al.\ (2000, 2002), but it was not expected for 
VLA~1 since this latter source traces an \UCHII\ region.  We will call VLA~1$^\star$ the VLBA source
detected here toward VLA~1, and discuss its nature further in the upcoming section.

\begin{figure}[!hb]
\begin{center}
\begin{tabular}{c}
\includegraphics[width=.87\linewidth,trim= 20 10 190 10, clip]{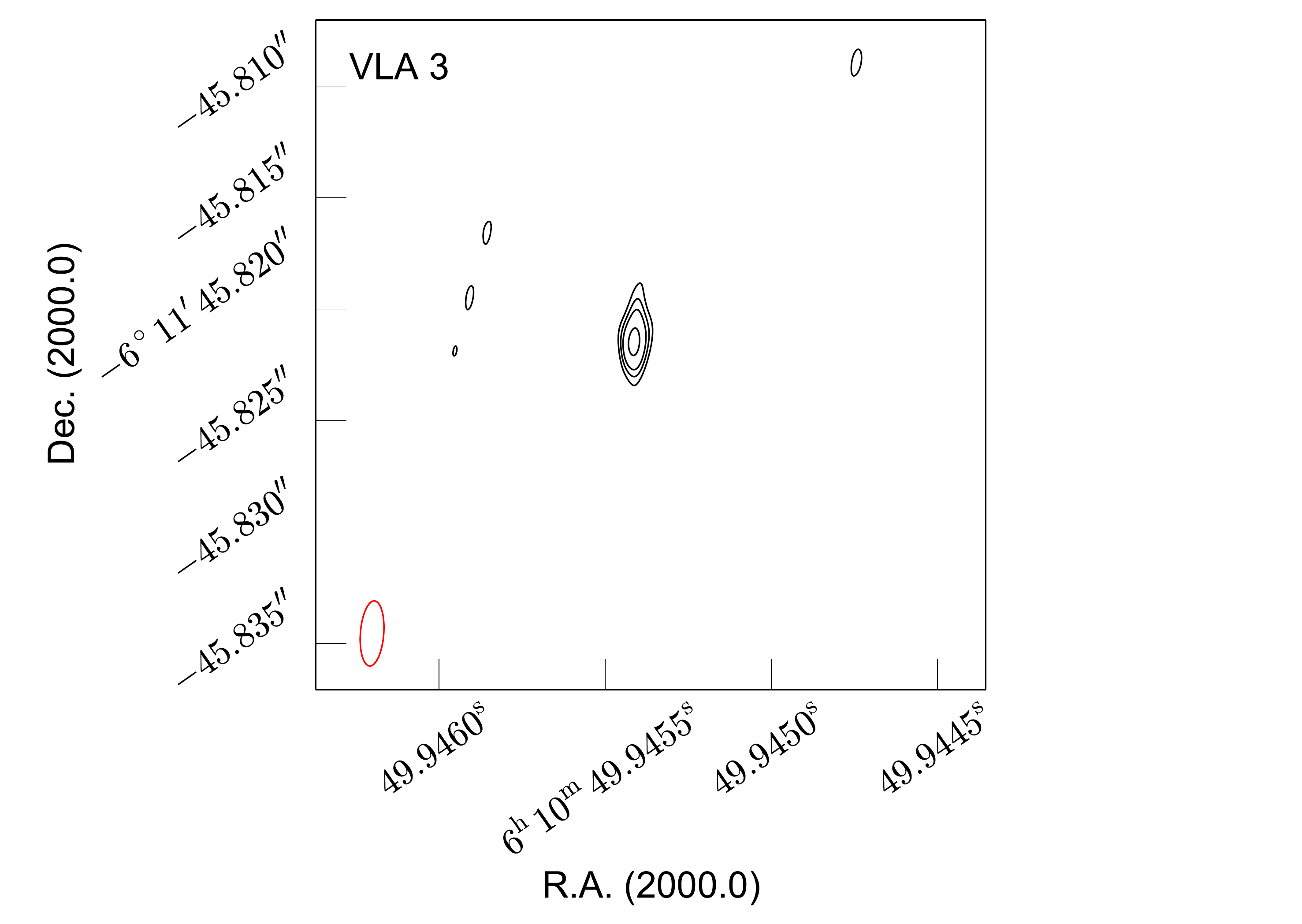}\\
\includegraphics[width=.87\linewidth,trim= 20 10 195 10, clip]{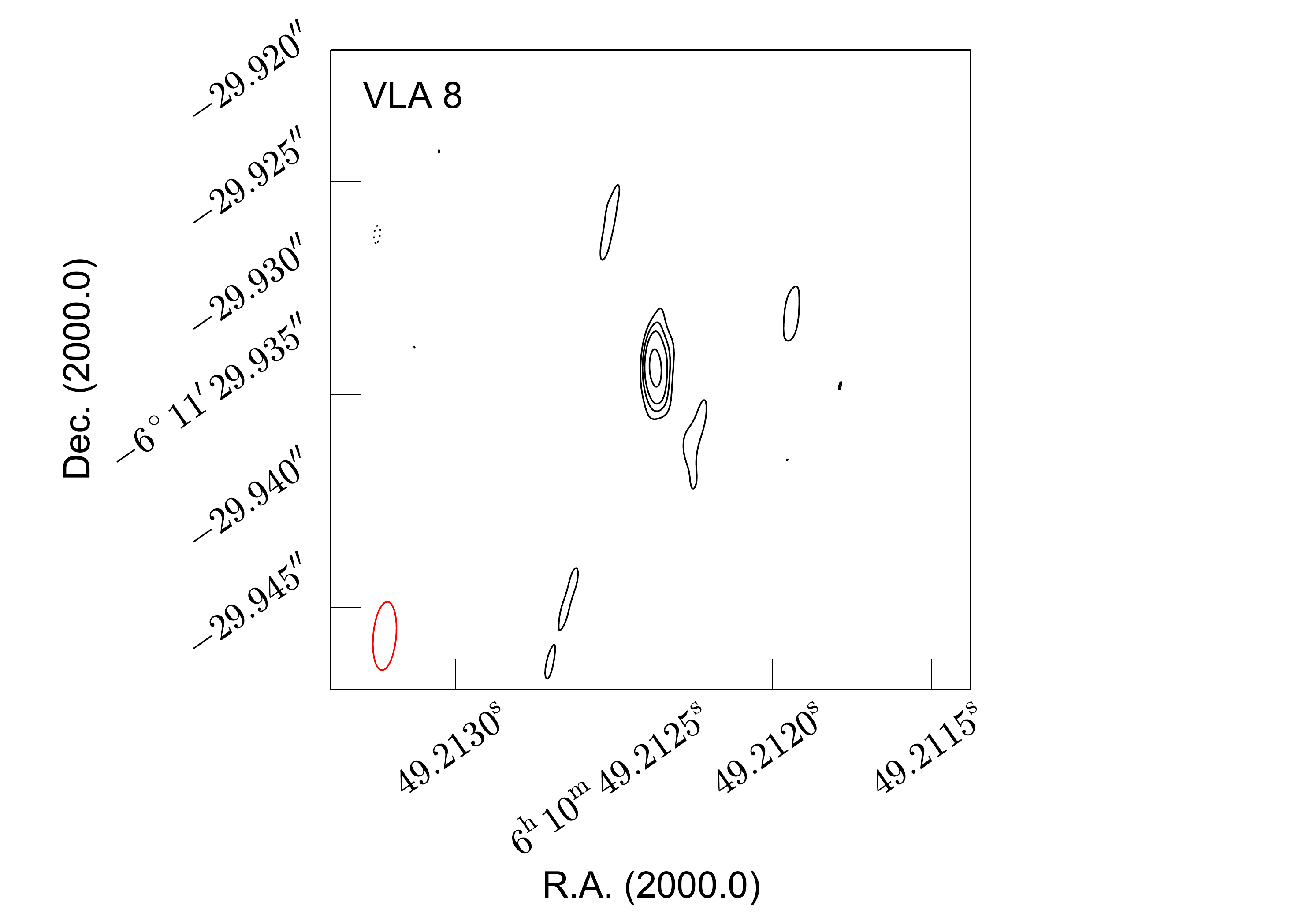}\\
\includegraphics[width=.87\linewidth,trim= 20 10 195 10, clip]{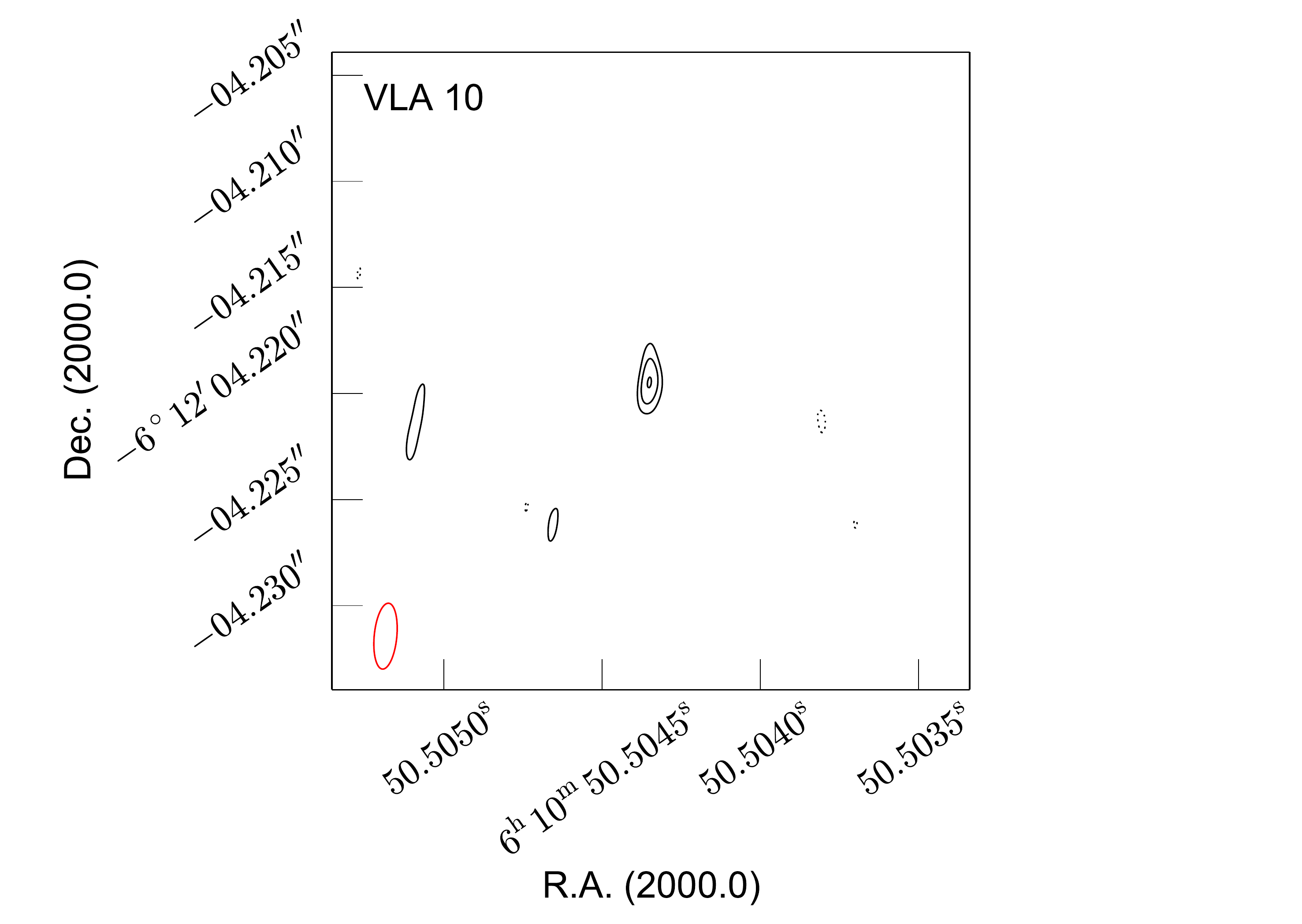}
\end{tabular}
\end{center}
\caption{Similar as Figure \ref{fig:VLA1det} but for the detection of VLA~3, VLA~8
and VLA~10. The contour levels are -3, 3, 4.5, 6 and 9
times the noise level.}
\label{fig:VLA3det}
\end{figure}

\section{Discussion}

\subsection{The distance to Monoceros}

\begin{figure*}[!ht]
\begin{center}
\includegraphics[width=.70\linewidth,trim= 0 10 10 30, clip]{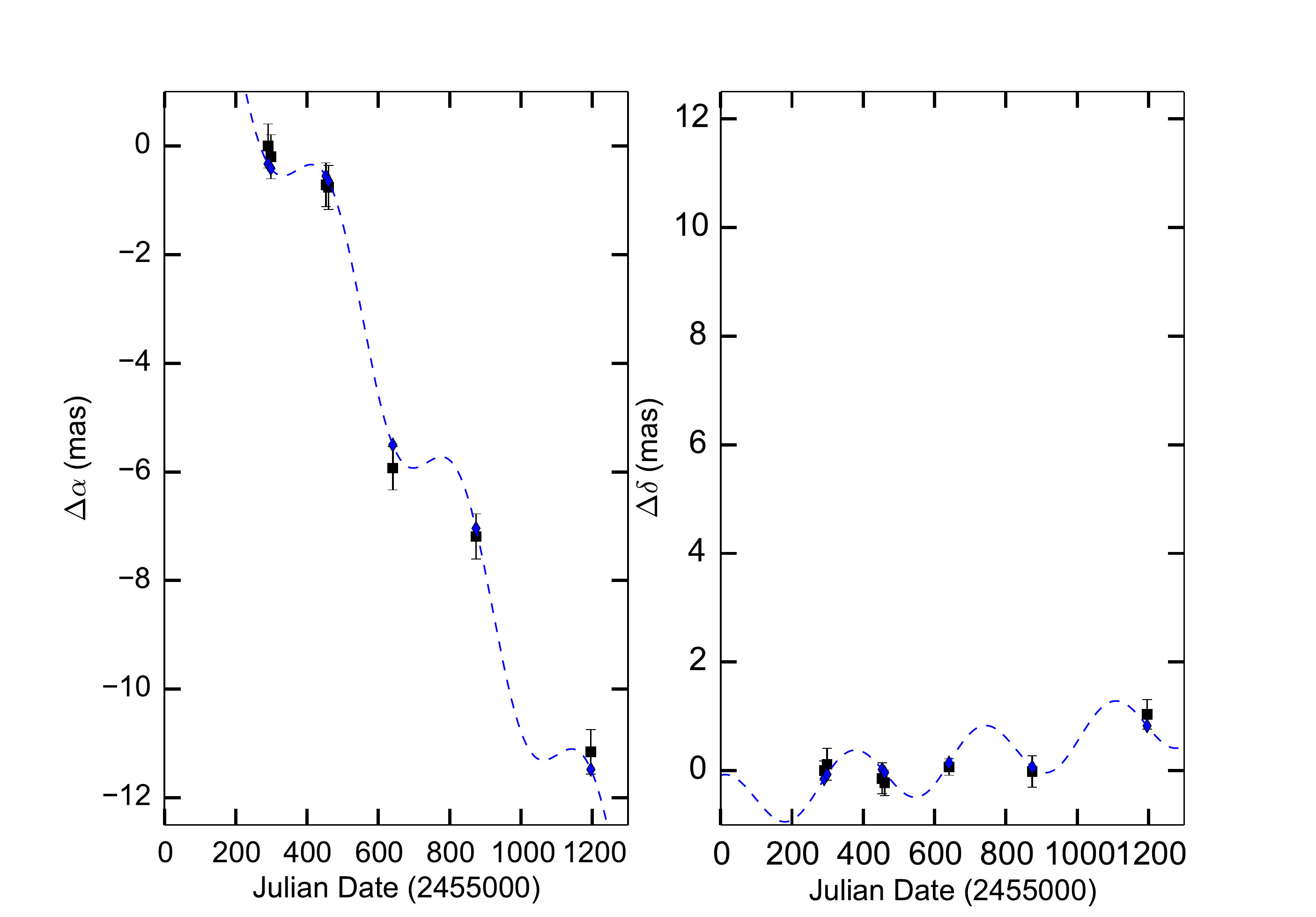}\\
\includegraphics[width=.70\linewidth,trim= 0 10 10 30, clip]{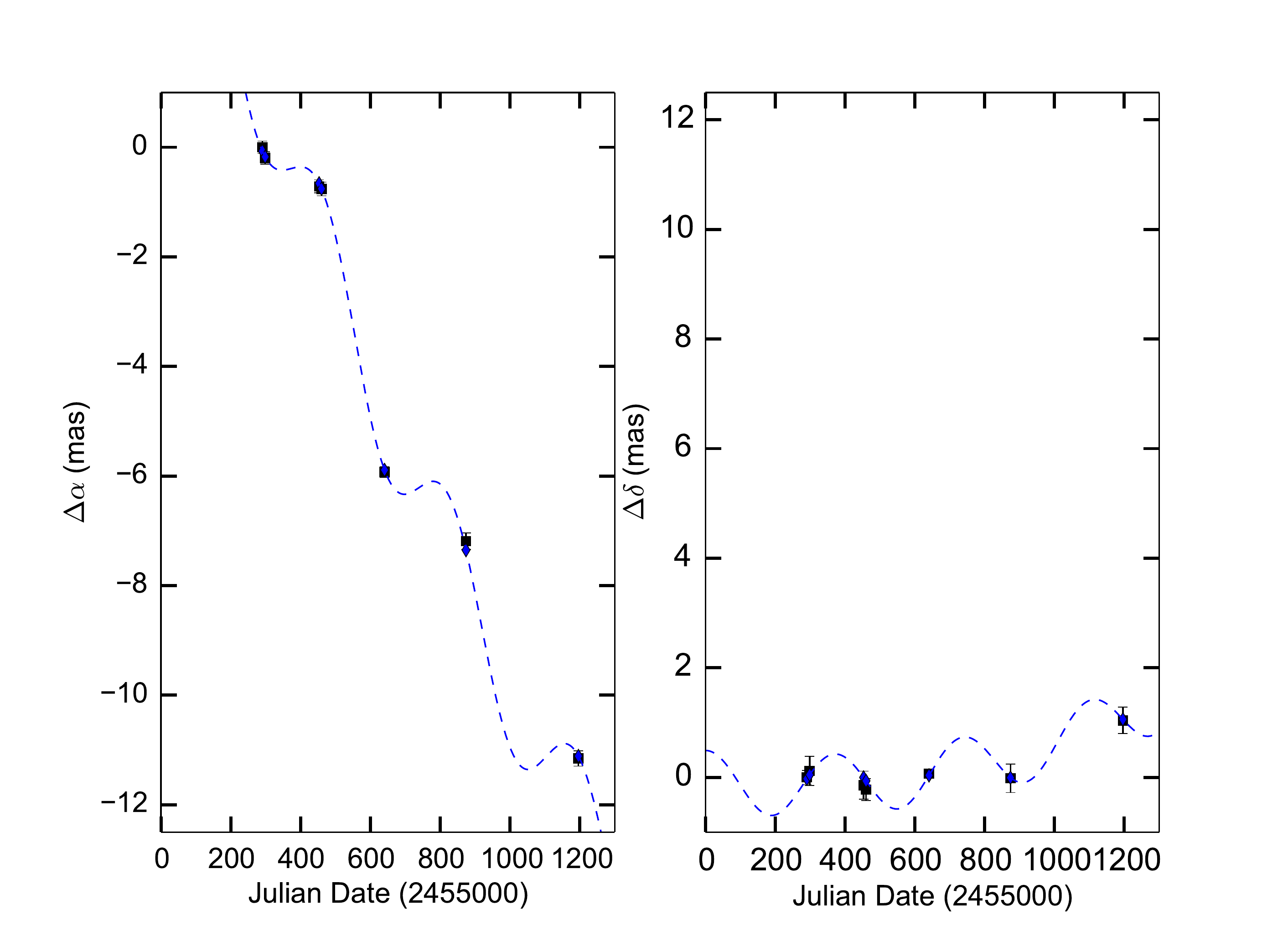}
\end{center}
\caption{The observed VLA~1$^\star$ positions for right ascension  (left panels) and declination (right panels)
are plotted as black filled squares. 
The dotted curves show the best fit to the trajectory of the star as a combination 
of the trigonometric parallax and uniform proper motion (top panels) and also including uniform
acceleration (bottom panels). Blue filled squares  are
the expected position according to the fit. }
\label{fig:pi}
\end{figure*}

The positions of VLA~1$^\star$ measured in our seven VLBA detections can be modeled as a combination of a trigonometric parallax ($\varpi$) and proper motion ($\mu$)  following e.g.\ Loinard et al.\ (2007). The barycentric coordinates of the Earth appropriate for each observation were calculated using the NOVAS routines distributed by 
the US Naval Observatory. The reference epoch was taken at { JD 2455743.22 $\equiv$ J2011.49}, the mean epoch of our detections. The best fit to the data assuming a uniform proper motion (Figure \ref{fig:pi}, top) yields the following astrometric elements:

\begin{eqnarray*}
\alpha_{J2011.49}             & = & 06^{{\rm h}}10^{{\rm m}}50\rlap.{^{\rm s}}591549 \pm 0\rlap.{^{\rm s}}000012\\
\delta_{J2011.49}             & = & -06^{\circ}11^{'}50\rlap.{''}38412 \pm 0\rlap.{''}00010\\
\mu_\alpha \cos{\delta}   & = & -5.35 \pm 0.21\ {\rm mas\ yr}^{-1}\\
\mu_\delta                       & = & 0.45 \pm 0.12\ {\rm mas\ yr}^{-1}\\
\varpi                               & = & 1.10 \pm 0.18\ {\rm mas}.
\end{eqnarray*}

The post-fit r.m.s. values, however, are fairly large: 0.27 and 0.16 mas in right ascension and declination, respectively. Indeed, systematic errors of 0.40 mas and 0.13 mas (in right ascension and declination, respectively) had to be added in quadrature to the uncertainties delivered by JMFIT to obtain a reduced $\chi^2$ of 1. 

The situation can be very significantly improved if the proper motion is allowed to be (uniformly) accelerated. The best fit, in this case (Figure \ref{fig:pi}, bottom) yields:

\begin{eqnarray*}
\alpha_{J2011.49}             & = & 06^{{\rm h}}10^{{\rm m}}50\rlap.{^{\rm s}}591523 \pm 0\rlap.{^{\rm s}}000005\\
\delta_{J2011.49}             & = & -06^{\circ}11^{'}50\rlap.{''}38422 \pm 0\rlap.{''}00009\\
\mu_\alpha \cos{\delta}   & = & -5.32 \pm 0.07\ {\rm mas\ yr}^{-1}\\
\mu_\delta                       & = & 0.50 \pm 0.10\ {\rm mas\ yr}^{-1}\\
a_\alpha \cos{\delta}   & = & 0.92 \pm 0.11\ {\rm mas\ yr}^{-2}\\
a_\delta                       & = & 0.37 \pm 0.14\ {\rm mas\ yr}^{-2}\\
\varpi                               & = & 1.12 \pm 0.05\ {\rm mas}.
\end{eqnarray*}

From the comparison with the non-accelerated case, it can be seen that the situation in 
declination remains nearly unchanged, with an acceleration component in that direction 
consistent with 0 within 2.5$\sigma$. In right ascension, however, the fit is significantly 
improved, with an acceleration component detected at nearly 9$\sigma$ and an improvement 
on the final parallax accuracy by a factor of 3. This improvement also results in the fact 
that to obtain a reduced $\chi^2$ of 1, no systematic errors at all had to be added to those 
delivered by JMFIT in declination. In right ascension, only 0.09 mas had to be added (compared
with 0.33 mas when no acceleration is allowed). Note that the best-fit value of the parallax
remains, nevertheless, fully consistent with the non-accelerated case.

The existence of acceleration in the proper motion of VLA~1$^\star$ suggests that it belongs
to a multiple system. The acceleration vector does not appear to point toward any of the 
known sources associated with VLA~1. In addition, the amplitude of the acceleration appears 
to be fairly substantial (nearly 1 mas yr$^{-2}$, compared with a proper motion of about 
5 mas yr$^{-1}$ along that same direction) so the companion must be very nearby. We are most 
likely dealing with a very compact binary stellar system. 

\begin{figure}[ht!]
\begin{center}
\includegraphics[width=1.0\linewidth]{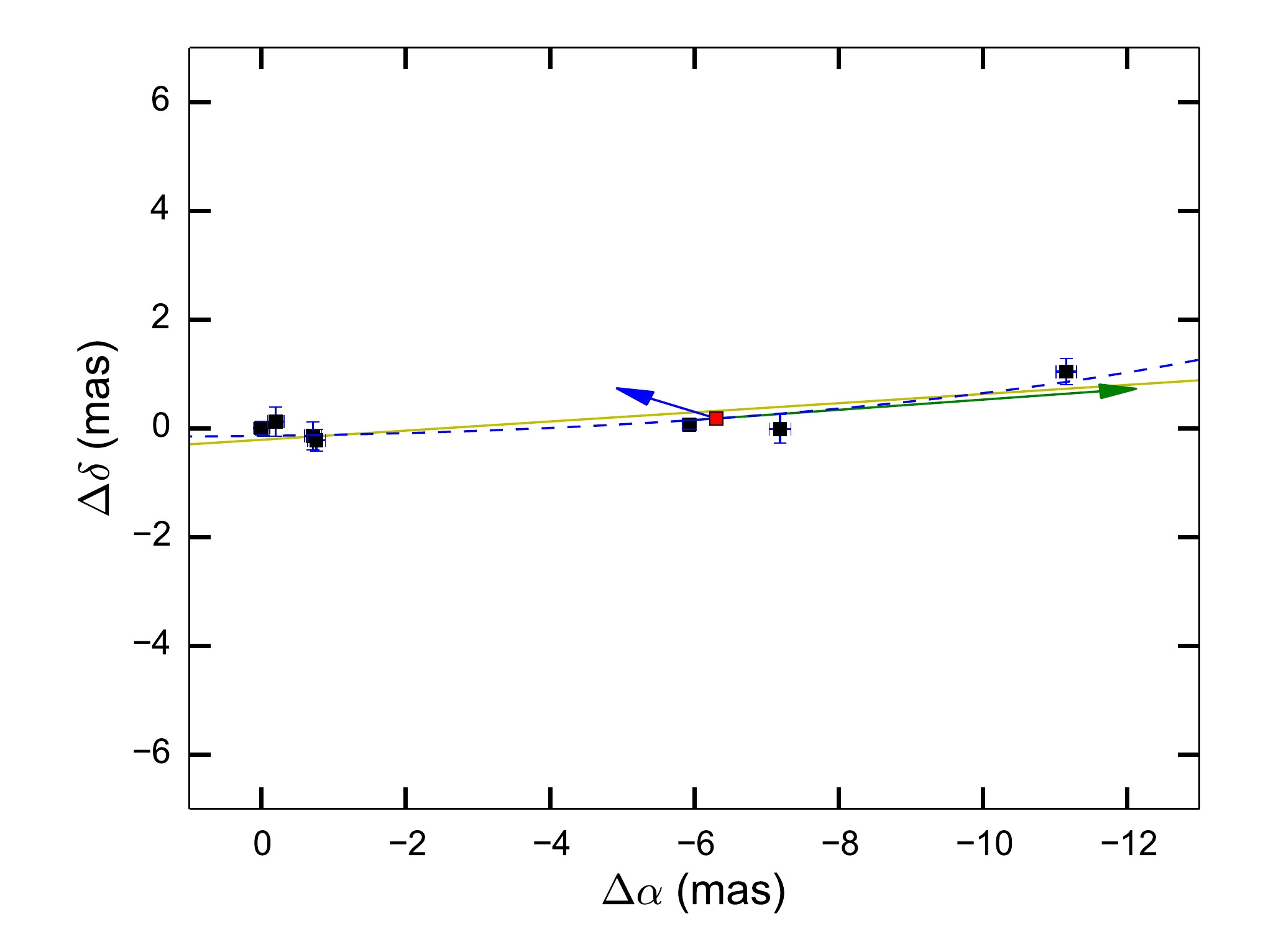}
\end{center}
\caption{The observed VLA~1$^\star$ positions, relative to the first detected epoch, with correction due to the trigonometric parallax. 
The position of VLA~1$^\star$ at the reference epoch is plotted as a red square.
The dotted curve shows the best fit after removing the  movement of the trigonometric parallax. 
The proper motion and acceleration vectors are shown as green and blue arrows, respectively. The yellow line
shows the expected movement of the star from the linear fit, without acceleration, after removing 
the trigonometric parallax.}
\label{fig:pSky}
\end{figure}

Since the acceleration can be approximated with a constant value over the
interval of 2.5 years spanned by
the observations, we can conservatively conclude that the period of the binary
must be significantly longer that 2.5 years { (e.g., see Figure \ref{fig:pSky})}. 
If we assume that the binary system is formed by two stars each with
one solar mass and that the
orbit is circular and face-on, we will get the largest acceleration possible
in the plane of the sky.
This gives an upper limit to the separation for the components of the binary
of 6.5 AU. In turn, this gives an upper limit to the period of 12.2 years.
We conclude that our results are consistent with a binary system with these parameters.
{ Since acceleration is not constant along a Keplerian orbit, future observations 
ought to be inconsistent with our constant acceleration assumption. In particular, 
observations obtained in the next few years should, if fitted independently of the
data presented here, yield a different acceleration vector.}

The best fit to our data implies a parallax $\varpi$ = 1.12 $\pm$ 0.05 mas, corresponding to a distance 
$d$ = 893$^{+42}_{-40}$ pc. This result is, within errors, in agreement with the distance of 830 $\pm$ 50 pc obtained by 
Herbst \& Racine (1976), and the kinematic distance of Rodr\'{\i}guez et al. (1980). It should be emphasized, however, that 
the previous measurements were indirect and, therefore, prone to possible systematic errors related with uncertain assumptions 
regarding the nature of the observed star, the properties of the interstellar medium, or the appropriate Galactic rotation curve. 
Our measurement, on the other hand, is purely geometric.

\subsection{Galactic kinematics of Monoceros}

It has been suggested that Monoceros and Orion might be linked by an expanding super-bubble
that may have triggered star-formation in both regions (Heiles 1998). From the distance to 
Monoceros reported above and its measured proper motion and radial velocity 
($v_{lsr}$ = +11 $\pm$ 1 km s$^{-1}$; see Rodr\'{\i}guez et al.\ 1980), one can derive the 
location $(x,y,z)$ and velocity vector $(u,v,w)$ of Monoceros in the Milky Way, and examine 
this issue from a kinematic point of view. We will express the positions and velocities in 
the commonly used rectangular reference frame centered on the location of the Sun, with the 
$(Ox)$ axis pointing toward the Galactic center, the $(Oy)$ axis perpendicular to $(Ox)$ and
pointing in the direction of Galactic rotation, and $(Oz)$ pointing toward the Galactic North
Pole (GNP) so as to make $(Oxyz)$ right-handed. In these coordinates, the position of Monoceros
appears to be $(x,y,z)$ = $(-846,-6,-283)$ pc: it is located almost exactly toward the Galactic
anticenter, but nearly 300 pc below the Galactic mid-plane. For reference, in the same system, 
Orion is located at $(-393, -3,-131)$ pc, nearly the same direction but at less than half the distance. 

\begin{figure*}[!ht]
\begin{center}
\includegraphics[width=1.0\linewidth,trim= 50 170 100 200, clip]{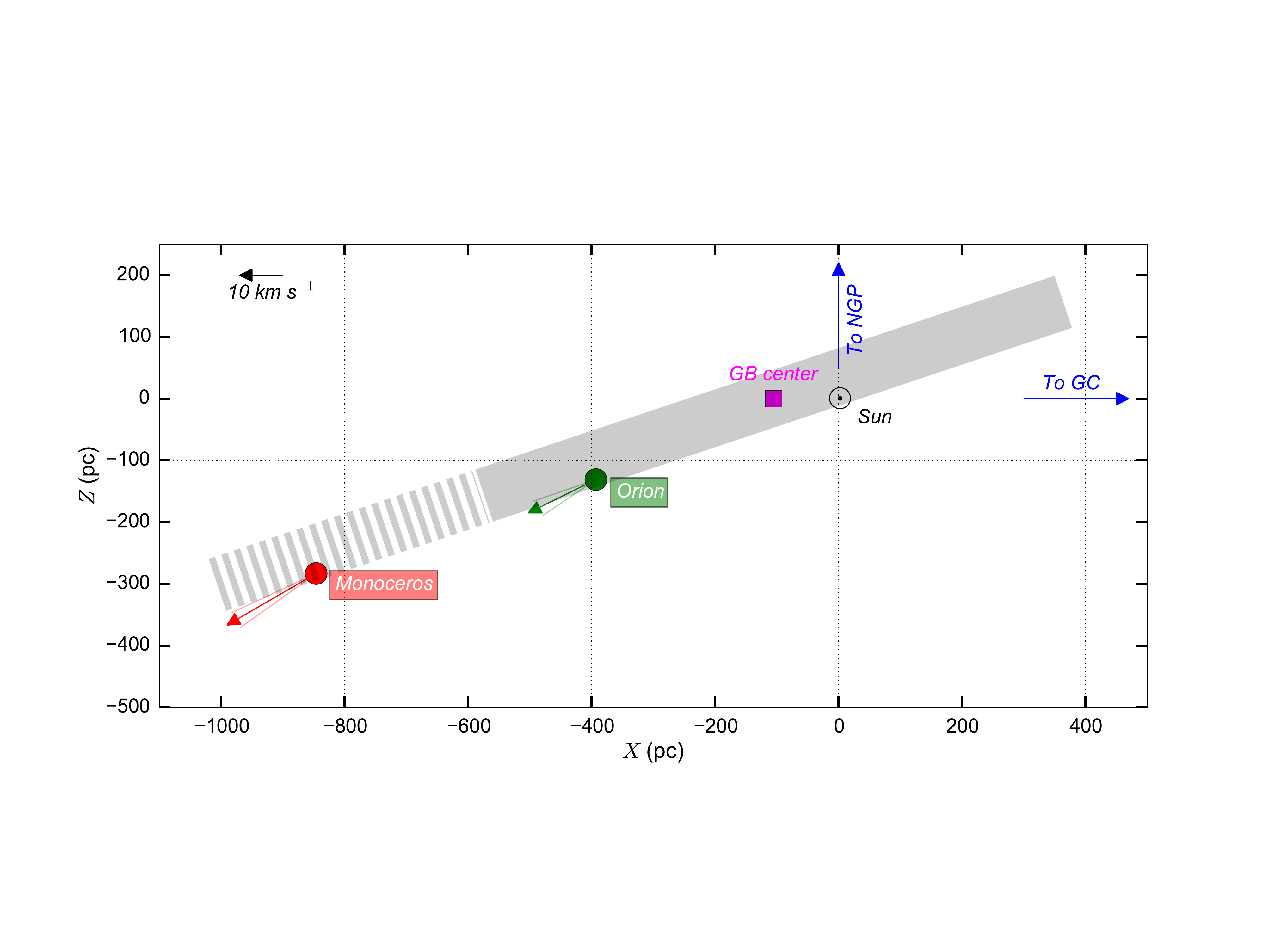}
\end{center}
\caption{Position and velocity vectors of Monoceros and Orion in the Galactic (X,Z) plane
with the reference frame centered on the Sun (see text for the definition of these
axes). The grey band indicates the location of Gould's Belt, with the center indicated
by a magenta square. Stripes indicate the extension of the Gould's Belt plane and
shows that the position and velocity vectors of Monoceros lie in this plane. Blue
arrows indicate the direction to the Galactic plane (GC) and to the northern Galactic
pole (NGP).}
\label{fig:GB_XZ}
\end{figure*}

The velocity vector of Monoceros deduced from our observations, and expressed in the same 
rectangular frame relative to the LSR is $(u,v,w)$ = $(-25.4,-10.3,-14.7)$ km s$^{-1}$ with 
formal uncertainties of order 2 km s$^{-1}$ along each direction. We note, however, that since 
this value is based on a single proper motion measurement, it could contain a significant 
systematic error if VLA~1$^\star$ happens to have a significant peculiar velocity relative 
to the rest of the Monoceros R2 cluster. In the same system, the velocity of Orion is
$(u,v,w)$ = $(-18.3,-15.6,-9.0)$ km s$^{-1}$. Thus, there does not appear to be an obvious 
relation between the relative velocities of the two regions. This does not discard a common origin, 
however, as dynamical evolution of expanding bubbles can be complex. As a final note, we would
like to point out that while Monoceros is not normally associated with Gould's Belt (e.g.\ P\"oppell 1997),
it does appear to be projected almost exactly along the plane defining Gould's Belt (Figure \ref{fig:GB_XZ}). Its 
velocity vector is also almost entirely contained within the plane defining Gould's Belt, 
and oriented mostly away from Gould's Belt center, as would be expected if it participated 
in the overall expansion of that structure.

\subsection{On the nature of VLA~1$^\star$}

\begin{figure}[!ht]
\begin{center}
\includegraphics[width=1.2\linewidth,trim= 90 0 110 10, clip]{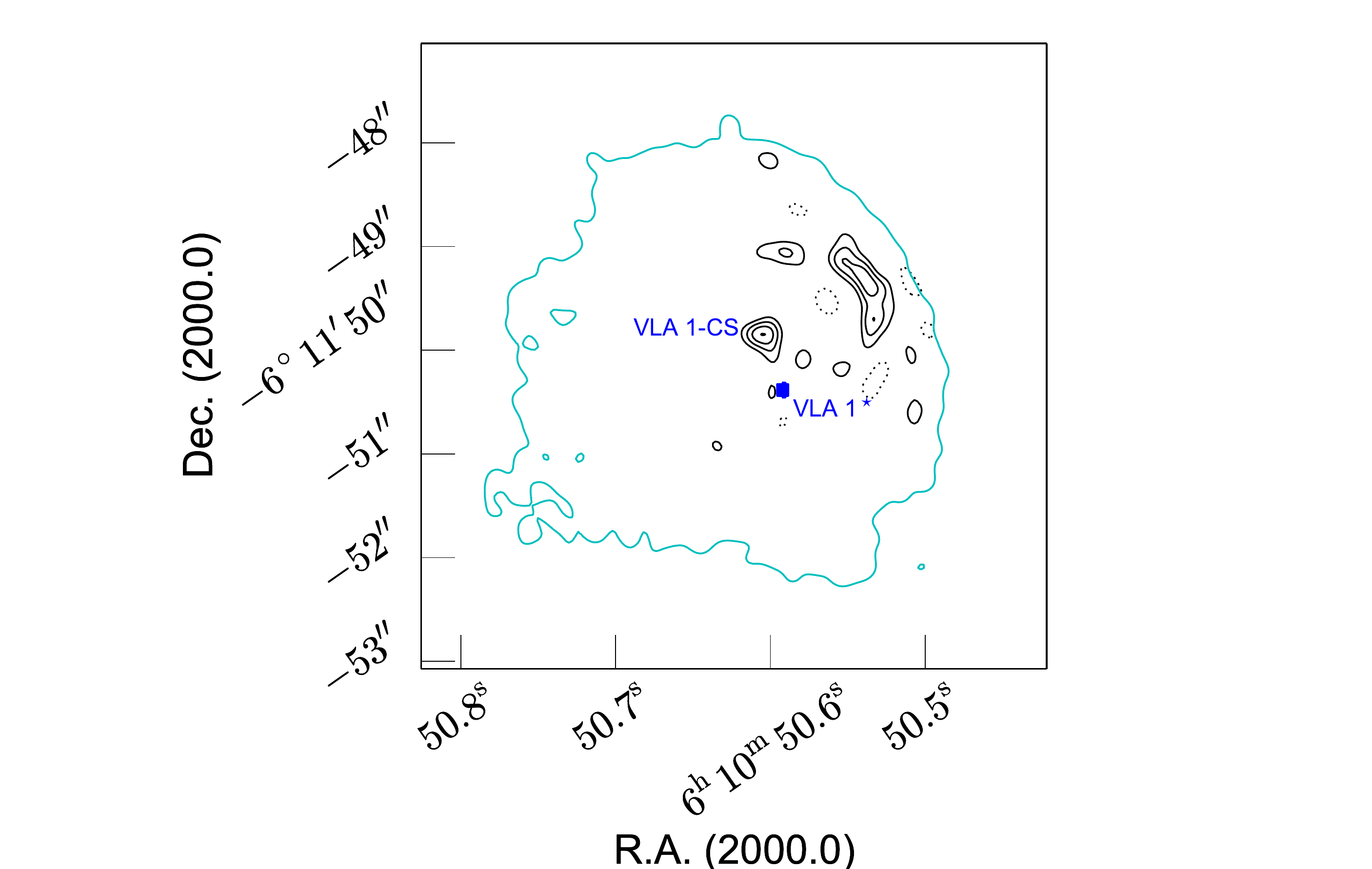}
\end{center}
\caption{The VLA~1 cometary UCH{\small II} region as seen in the VLA continuum map at 8.4 GHz,
as shown by G\'omez et al. (2000). The cyan line is the border of the 
extended radio emission. The black contours trace the compact radio emission when visibilities
shorter than 200k$\lambda$ are removed. The compact source studied by G\'omez et al.~(2000)
is labeled as VLA~1-CS. The blue square indicates the position of VLA 1$^\star$.}
\label{fig:hii}
\end{figure}

Our detection of a very compact non-thermal radio source close to peak position of the \UCHII\ region 
VLA~1 (G\'omez et al.\ 1998) is somewhat unexpected. Interestingly, this is not the first compact object
seen in projection toward VLA~1. G\'omez et al.\ (2000) used the VLA telescope to observe the GGD 12-15
region, and by removing the shortest baselines (i.e., removing the extended emission) they found a slightly
resolved source near the center of the extended emission associated with VLA~1 (see Figure \ref{fig:hii}).
This compact source, however, does not coincide in position with VLA~1$^\star$, as can also be seen in 
Figure \ref{fig:hii}. The separation between both sources is of 0\rlap{$''$}\,.55 which is significantly larger
than the position errors of the VLA images ($\sim$0\rlap{$''$}\,.1; Gomez et al. 2000). The proper motion 
of VLA~1$^\star$ over the 15 years separating the observations reported by G\'omez et al.\ (2000) and 
our VLBA observations is expected to have been only \msec{0}{08}, also insufficient to explain the 
separation between VLA~1$^\star$ and the compact source reported by them.
In addition, while the brightness temperature of the compact source detected by
G\'omez et al.\ (2000) is $\sim$200 K, that of VLA~1$^\star$ is larger than 10$^6$ K. This further confirms
that the two sources have a different nature.

Compact radio sources have been founded toward other \UCHII\ regions: NGC 6334A (Carral et al.\ 2002;
Rodr\'{\i}guez et al.\ 2014), NGC 6334E (Carral et al.\ 2002), and W3(OH) (Dzib et al.\ 2013, 2014). 
The source detected in W3(OH) is also thermal in nature, but those detected in NGC 6334A and NGC 6334E 
have been suggested to be non-thermal sources (Carral et al.\ 2002; Rodr\'{\i}guez et al.\ 2014). However, 
this is the first time that a compact source projected toward an \UCHII\ region is imaged with the VLBI technique.

Massive stars are expected to be fully radiative. Thus, they are not expected to produce magnetically 
active coronas. On the other hand they can produce non-thermal radio emission in the strong shocks produced by
the interaction of 
their jets with the surrounding medium (e.g., Carrasco-Gonz\'alez et al.\ 2010) and in the wind-wind 
collision regions of two massive stars (e.g., Ortiz-Le\'on et al.\ 2011). The only well documented cases 
of jet-medium shocks show that they produce emission that is extended to sizes of around $1''$ level (e.g.,
Carrasco-Gonz\'alez et al.\ 2010; Wilner et al. 1999), that will be resolved out in VLBI observations. 
On the other hand, wind-wind collision regions have been imaged with VLBI
and appear as resolved bow shock structures (Dougherty et al.\ 2005; Ortiz-Le\'on et al.\ 2011). At the
distance of Monoceros such structures should be resolved if present, but this is not the case of the source
reported here that shows, instead, a very compact structure in the seven detected epochs. The brightness temperature, 
size and variability of VLA~1$^\star$ is similar to those detected 
for magnetically active intermediate- and low-mass YSOs (e.g., Loinard et al.\ 
2008, Dzib et al.\ 2010). As a consequence, we suggest that this source  is a magnetically active star 
(this explains, a posteriori, our choice of name, VLA~1$^\star$, for this source). With the data presented 
here, we cannot constrain the properties of VLA~1$^\star$ further. In particular, it is unclear whether 
VLA~1$^\star$ is physically associated with the \UCHII\ region VLA~1, or if it is located behind or in front
of it. Also, the mass of the star associated with VLA~1$^\star$ remains entirely unconstrained, as are the
properties of the putative multiple system to which it belongs.

\subsection{On the nature of the remaining compact radio sources}

Most of the radio compact sources in GGD 12-15 were associated with magnetically active T Tauri stars
by G\'omez et al.\ (2002), due to their variability and spectral indices. The only exception was VLA~7
whose steady flux density and positive spectral index suggested a thermal radio emission produced by 
stellar winds. With the new radio observations, the non-thermal nature is confirmed for VLA 3, VLA 8, 
and VLA 10. 

Adopting 5$\sigma$ upper limits, the remaining five sources were not detected at levels of 90 $\mu$Jy. 
This flux level is comparable to (or, in some cases, significantly lower than) the flux density reported
by G\'omez et al. (2002) for these sources. An explanation is that their flux levels during all the 
observations were lower than our threshold detection limit. This is feasible, since also the flux density
of VLA 3, VLA 8 and VLA 10 are below our threshold limit in almost all the epochs. Another explanation is
that they are resolved out in our high resolution images. As a rough attempt to detect extended non-thermal
radio emission from these sources, we produced images by only using the shortest baselines in the array 
($<$80,000 k$\lambda$). No detection was obtained.

Finally, we should mention that our observations at 8.4 GHz are not ideal for these sources,
given their negative spectral indices. New VLBI observations at longer wavelengths would help to clarify
the nature of their radio emission. This would be possible using the new sensitive C-band (6 GHz)
receivers installed on the VLBA antennas. 

\section{Conclusions}

In this paper, we reported on VLBA observations of the cluster of compact radio sources associated
with the \UCHII\ region in GGD 12-15, also known as VLA~1, reported  by G\'omez et al.\ (2000, 2002).
We detected a total of four sources. The sources VLA~3, VLA~8, and VLA~10 were detected only in single
epochs, while a source projected inside VLA~1 was detected in seven different epochs. Given its radio 
properties, this source is likely to be a magnetically active star, that we call VLA~1$^\star$. Using 
the measured positions of VLA~1$^\star$ in the seven detected epochs, we derive its proper motion and 
trigonometric parallax, $\varpi$. The best fit yields $\varpi$ = 1.12$\,\pm\,$0.05 mas, that
corresponds to a distance $d$ = 893$^{+42}_{-40}$ pc. We argue that this is the most reliable distance 
measurement to the Monoceros R2 region to date.

\acknowledgments
G.N.O.-L, L.L., and L.F.R.\  acknowledge the financial support of DGAPA, UNAM, and CONACyT, M\'exico. 
L.L.\ and G.N.O.-L are indebted to the Alexander von Humboldt Stiftung for financial support. The National
Radio Astronomy Observatory is operated by Associated Universities Inc. under cooperative agreement with 
the National Science Foundation.

\end{document}